\documentclass[a4paper]{jpconf}
\usepackage{graphicx}
\begin{document}
\title{Transition from Galactic to Extra-Galactic Cosmic Rays}

\author{Roberto Aloisio}

\address{INFN - Laboratori Nazionali del Gran Sasso, 
I--67010 Assergi (AQ), Italy}

\ead{roberto.aloisio@lngs.infn.it}

\begin{abstract}
In this paper we review the main features of the observed Cosmic Rays 
spectrum in the energy range $10^{17} {\rm eV}~\div~10^{20} {\rm eV}$.
We present a theoretical model that explains the main observed features 
of the spectrum, namely the second Knee and Dip, and implies a transition 
from Galactic to Extra-Galactic cosmic rays at energy $E\simeq 10^{18}$ eV, 
with a proton dominated Extra-Galactic spectrum.
\end{abstract}

\section{Introduction}
The CR spectrum observed on Earth extends over many orders of magnitude
from GeV energy up to energies larger than $10^{20}$ eV. This spectrum 
is a steeply falling power law with almost no structure apart from four 
still not well understood features. These features can be summarized as 
follows:
(i) at energy of about $4\times 10^{15}$ eV there is a change in the
spectral index, called the {\it Knee}; 
(ii) at energy around $4\times 10^{17}$ eV, the spectral index changes 
again with a steeper spectrum, this feature is called the {\it 2nd Knee}; 
(iii) in the energy range $10^{18} {\rm eV}~\div~8\times 10^{19} {\rm eV}$ 
the spectrum presents a smooth suppression called the 
{\it Dip}, with a flattening around $E\simeq 10^{19}$ eV called the 
{\it Ankle}; 
(iv) finally, the debated Greisen-Zatsepin-Kuzmin (GZK) feature \cite{GZK} 
could be present in the spectrum starting from $8\times 10^{19}$ eV.
Leaving aside the GZK, whose presence will be soon clarified by the 
Auger detector \cite{detectors}, the other features can be well recognized 
in the experimental data of all the CR detectors. The two knees structure 
as well as the presence of the Dip is firmly established by different detectors
\cite{detectors}. While, from an experimental point of view, the presence 
of the 2nd Knee and Dip is not questionable their theoretical interpretation 
is still under debate. A key ingredient in any theoretical model explaining 
the 2nd Knee and Dip is the CR chemical composition at the Dip energies, 
namely in the energy range $10^{18} {\rm eV}~\div~8\times 10^{19} {\rm eV}$. 
>From an experimental point of view this chemical composition is still 
not well understood, there are opposite indications that favor different 
scenarios: the HiRes, HiRes-MIA and Yakutsk detectors favor a proton 
dominated flux at energies $E>10^{18}$ eV \cite{detectors}, while, in the 
same energy range, Fly's Eye, Haverah Park and Akeno detectors indicate a 
mixed composition with protons and heavy nuclei (most probably Iron nuclei) 
\cite{detectors}. 
Another important piece of information comes from the energy range below 
the 2nd Knee, namely at $E<10^{17}$ eV. The origin of CR in this energy
range is clearly galactic and the Kascade data \cite{detectors}, that are 
in good agreement with all other measurements, show a gradual transition from 
light to heavy nuclei starting from the first knee. This behavior is well 
explained in the rigidity propagation/acceleration models (for a review see 
\cite{rig_rev}, and references therein) according to which every spectrum
of each galactic nuclear species presents a steepening at the energy 
$E_Z = Z E_p$, with $E_p=2.5\times 10^{15}$ eV associated to protons. 
In these models the first knee coincides with the energy at which the proton
spectrum shows a steepening and the subsequent behavior of the spectrum 
results from the convolution of the spectra of heavier nuclei, each 
characterized by a knee (steepening energy). According to this picture 
above the Iron knee $E_{Fe} = 6.5 \times 10^{16}$ eV the Galactic Cosmic 
Rays (GCR) spectrum should be suppressed.

The observed behavior of the all-particle spectrum at the Ankle energy
($E\simeq 10^{19}$ eV) was traditionally interpreted as the transition
from GCR to Extra-Galactic Cosmic Rays (EGCR) \cite{trad}. Nevertheless, 
this interpretation presents several problems as discussed in detail in 
\cite{preparation} mainly related to the difficulty in extending the GCR 
flux, thought to be dominated by heavy nuclei, up to very high energies 
($10^{19}$ eV). In this paper we briefly present an alternative model that 
places the transition GCR-EGCR at lower energies, namely around the 2nd Knee 
energy ($10^{18}$ eV). This model is based on an alternative explanation of 
the all-particle spectrum at the Dip energies. In the next section we will 
present the main features of our model.
 
\section{Transition from Galactic to Extra Galactic Cosmic Rays}

\begin{figure}[t]
\begin{minipage}{15pc}
\includegraphics[width=15pc]{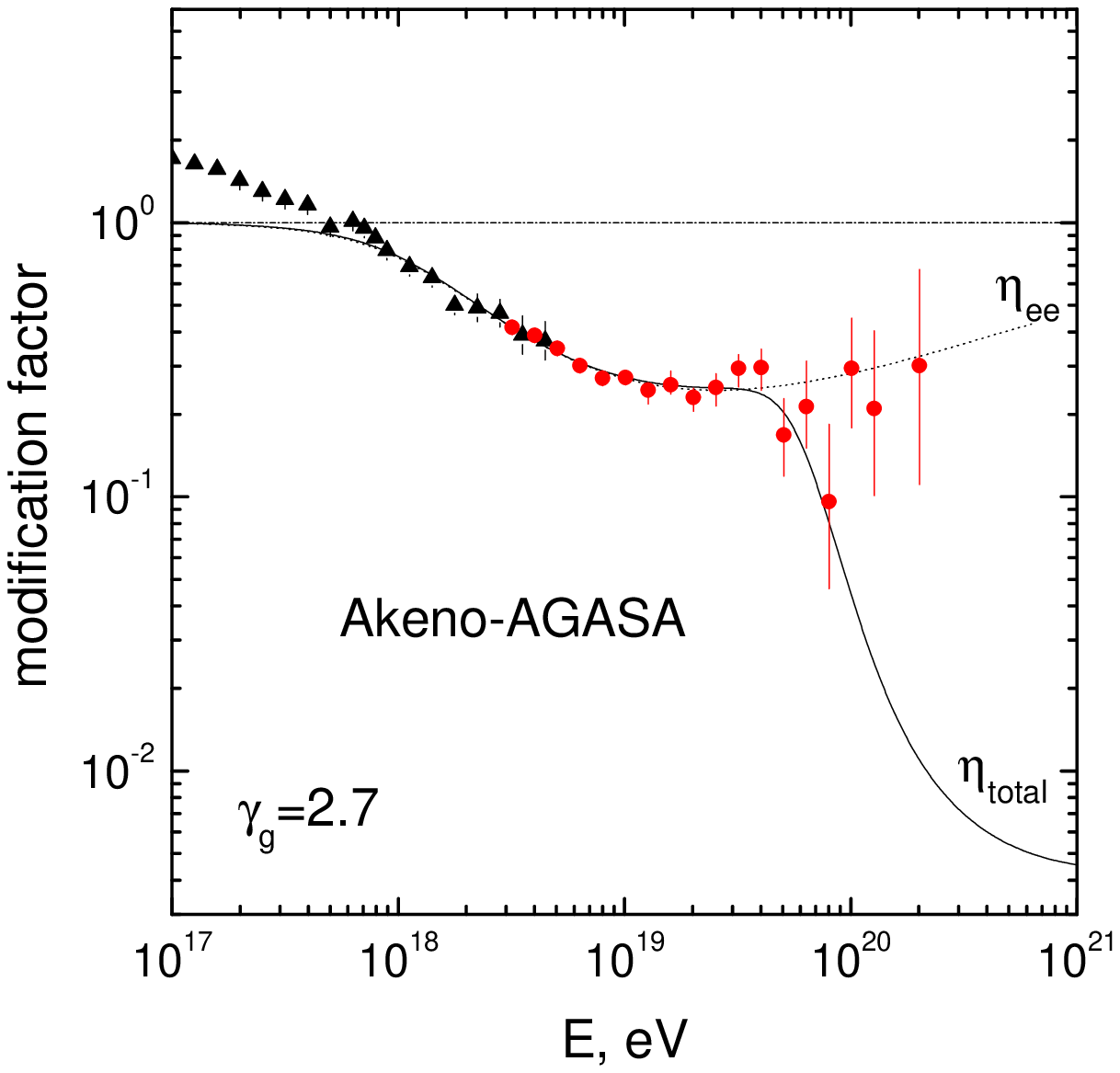}
\caption{\label{fig1} Modification factor as discussed in the text, 
compared with the AGASA data.}
\end{minipage}
\hspace{6pc}%
\begin{minipage}{17pc}
\includegraphics[width=17pc]{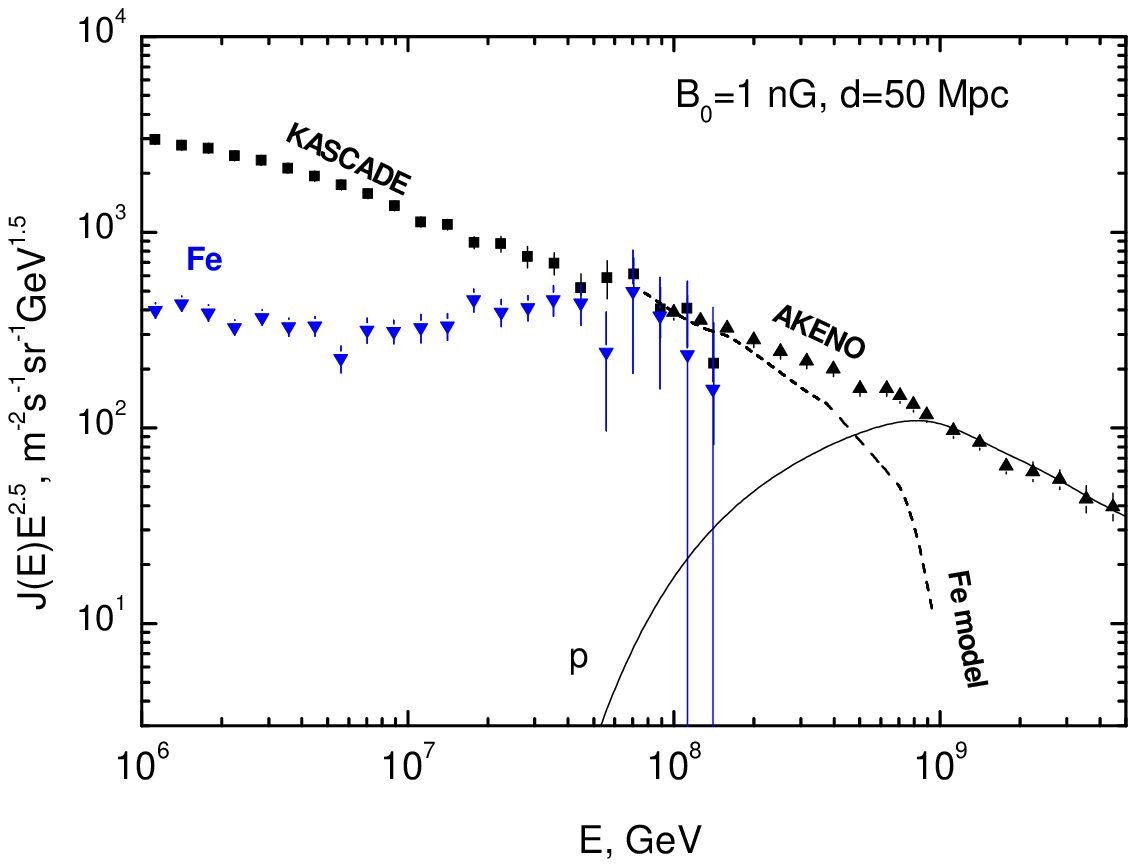}
\caption{\label{fig2} Model of transition GCR-EGCR as discussed in the text.}
\end{minipage} 
\end{figure}

Following \cite{Dip} (see also references therein) there are convincing 
evidences that the presence of the observed Dip, in contrast to the 
traditional interpretation, signals a proton dominated spectrum
at energies $E>10^{18}$ eV. As in \cite{Dip} we will use the formalism of the 
modification factor $\eta(E)$ defined as the ratio of the spectrum 
$J_p(E)$, with different channels of energy losses taken into account, 
and the unmodified spectrum $J_p^{unm}$, where only adiabatic energy losses 
(red-shift) are included: $\eta(E)=J_p(E)/J_p^{unm}(E)$. In figure \ref{fig1}
we report three different modification factors: the total one $\eta_{tot}$,
in which $J_p(E)$ is the proton spectrum with all channels of energy losses
taken into account, $\eta_{ee}$ in which $J_p(E)$ includes only adiabatic 
and pair production energy losses and the observed modification factor
$\eta_{obs}$ that uses $J_{obs}(E)$ as in the AGASA data \cite{detectors}. 
The injection spectrum used in figure \ref{fig1} is a single power law, 
with spectral index $\gamma_g=2.7$, and a source total emissivity per unit 
of comoving volume ${\cal L}_0=1.5 \times 10^{44}$ erg/Mpc$^3$s. This value
of the total emissivity is very high because of our assumption
of a single power law at injection (with $\gamma_g=2.7$) from
$E_{min}=1$ GeV up to $E_{max}=10^{21}$ eV. To reduce the required emissivity
one can assume that the acceleration mechanism works only from a somewhat 
higher minimum energy \cite{preparation}.
The behavior of the modification factor $\eta_{ee}$, as reported in figure 
\ref{fig1}, reaches a very good agreement with experimental data (as reported
in \cite{Dip} $\chi^2$/d.o.f.=1.12). This is a strong indication that the  
Dip can be explained as the effect of the pair production process in the 
interaction of UHE protons with the CMB radiation. From the behavior of the 
observed modification factor $\eta_{obs}$ it follows that at energies around 
the 2nd Knee this quantity, which is bound to be $\leq 1$ by definition, 
exceeds unity (see figure \ref{fig1}). This is the indication that a new 
component, of different origin, is dominating the spectrum. We interpret 
this new component as due to GCR. In this context the 2nd Knee energy can be 
interpreted as the energy where, going from high to low energy, the GCR 
component starts to enter the all-particle spectrum, in agreement with the 
Kascade data \cite{detectors} and with the rigidity acceleration/propagation 
models \cite{rig_rev}. The details of the transition between GCR and EGCR, 
namely the exact mixed composition in the energy range 
$10^{17} {\rm eV}~\div~10^{18} {\rm eV}$, depends, on the GCR side, on the 
specific model of propagation/acceleration chosen \cite{rig_rev} and, on the 
EGCR side, on the details of the proton propagation. The EGCR proton 
propagation at energy between $10^{17}$ eV and $10^{18}$ eV is affected 
mainly by the presence of an Intergalactic Magnetic Field (IMF), that 
provides a steepening of the EGCR flux at energies 
$E\leq 10^{18}$ \cite{anti-GZK}. The energy scale of this steepening has a 
universal nature, being the energy scale at which adiabatic and pair 
production energy losses occur at the same rate \cite{anti-GZK}. On the other
hand, the behavior of the spectrum at lower energies (i.e. $10^{17}~{\rm eV}
\le E \le 10^{18}~{\rm eV}$) is related also to the magnetic 
field strength chosen \cite{anti-GZK}. The transition scenario outlined is
described by figure \ref{fig2}, with the all-particle spectrum of Kascade
and Akeno \cite{detectors} and the EGCR spectrum obtained in the Bohm 
diffusive approximation with an IMF of $1$ nG and a source minimal distance 
from the observer of $50$ Mpc. In figure \ref{fig2} the dashed line 
represents the GCR component and is obtained as a result of subtracting 
the EGCR spectrum from the observed all-particle spectrum.

Summarizing, the firm experimental detection of the 2nd Knee and Dip founds 
a compelling explanation in terms of composition and origin of CR 
particles. In this context we propose the following model: (i) The 
detection of the Dip implies a proton EGCR component that dominates the 
CR spectrum starting from the 2nd Knee energy $E\simeq 10^{18}$ eV; 
(ii) the 2nd Knee energy assumes a universal meaning being related only to 
the protons energy losses mechanism; (iii) The expected transition from 
GCR to EGCR in the all particle spectrum, signaled by an observed 
modification factor $\eta_{obs}$ larger than one, sits between the 
detected Iron Knee $E_{26}=6.5\times 10^{16}$ eV and the 2nd Knee energy, 
the details of this transition being related, on the EGCR side, to the IMF. 
Finally, we can conclude stressing that a clear cut indication about the 
validity of the proposed models for the GCR-EGCR transition seems mainly 
related to a precise determination of the UHECR chemical composition at 
the Dip energies.

\vskip 0.5cm
\section*{Acknowledgments}
I am grateful to Venyamin Berezinsky, Pasquale Blasi, Askhat Gazizov and 
Svetlana Grigorieva with whom the present work was developed.

\end{document}